\documentstyle[aps]{revtex}

\begin{document}
\title{Wormholes, naked singularities and universes of ghost radiation}
\author{L\'{a}szl\'{o} \'{A}. Gergely}
\address{Astronomical Observatory and Department of Experimental Physics,\\
University of Szeged, D\'{o}m t\'{e}r 9, Szeged, H-6720 Hungary}
\maketitle

\begin{abstract}
Both the static and homogeneous metrics describing the spherically symmetric
gravitational field of a crossflow of incoming and outgoing null dust
streams are generalized for the case of the two-component ghost radiation.
Static solutions represent either naked singularities or the wormholes
recently found by Hayward. The critical value of the parameter separating 
the two possibilities is given. The wormhole is allowed to have positive 
mass. The homogeneous solutions are open universes. 
\end{abstract}

\bigskip

The static, spherically symmetric solution describing the gravitational
field of a crossflow of null dust streams \cite{GergelyND} was recently
extended by Hayward \cite{Hayward} for the case of two-component ghost
radiation, the resulting solution being interpreted as a wormhole
\footnote{For recent developments in the subject see \cite{Visser}, 
\cite{Hayward2} and for an earlier review \cite{Visser2}
}. The
original static solution \cite{GergelyND} however has a homogeneous
counterpart \cite{GergelyND2}-\cite{GergelyND3}, the similar generalization
of which is the principal aim of this letter. We also find the critical value 
of one of the metric parameters, which separates naked singularities from 
wormholes in the static case. We show, that despite the negative energy 
density of the ghost radiation, the wormholes can have positive mass.

{\bf The null dust solutions} can be concisely written as 
\begin{equation}
ds^{2}=\frac{ae^{L^{2}}}{cR}\left( dZ^{2}-2R^{2}dL^{2}\right) +R^{2}d\Omega
^{2}\ ,  \label{metric2nd}
\end{equation}%
where $c=-1$ refers to the static case (naked singularities) and $c=1$ to 
the homogeneous case (closed universe).
Providing $R>0$, the metric has the signature $\left( c,-c,+,+\right) $.
Accordingly , $Z$ is time coordinate in the static case and radial
coordinate in the homogeneous case while $L$ is time coordinate in the
homogeneous case and radial coordinate in the static case. The parameter's
notation was changed to $a=1/C>0$ for easier comparison with Ref. \cite%
{Hayward}. A second parameter $B$ is contained in the metric function $R$: 
\begin{equation}
cR=a\left( e^{L^{2}}-2L\Phi _{B}\right) ,\qquad \Phi
_{B}=B+\int^{L}e^{x^{2}}dx\ .  \label{cRL}
\end{equation}%
Both metrics have a true singularity at $R=0.$ (${\cal R}_{ab}{\cal R}%
^{ab}=2/a^{2}e^{2L^{2}}R^{2}$ and the Kretschmann scalar ${\cal R}_{abcd}%
{\cal R}^{abcd}$ contains $R^{6}$ in the denominator.) The energy-momentum
tensor is 
\begin{equation}
T_{ab}=\frac{\beta }{8\pi GR^{2}}\left( u_{a}^{+}u_{b}^{+}\pm
u_{a}^{-}u_{b}^{-}\right)  \label{enmom}
\end{equation}%
where $u^{\pm }$ are duals to the relatively normalized ($g\left(
u_{+},u_{-}\right) =-1)$ propagation null vectors 
\begin{equation}
\sqrt{2}u_{\pm }=\frac{1}{\sqrt{cg_{00}}}\frac{\partial }{\partial Z}\pm 
\frac{1}{\sqrt{-cg_{11}}}\frac{\partial }{\partial L}\ .  \label{nullvect}
\end{equation}%
From the Einstein equation the function $\beta $ is found positive, 
irrespective of the value of $c$:%
\begin{equation}
\beta =\frac{R}{ae^{L^{2}}}\ .  \label{beta}
\end{equation}%
The source can be equally interpreted as an anisotropic fluid with no
tangential pressures and both the energy density and radial pressure equal
to $\beta /8\pi GR^{2}=1/8\pi Gae^{L^{2}}R$ (these also became infinite at $R=0$).
As a third interpretation, it represents a massless scalar field in the 
two dimensional sector obtained by spherically symmetric reduction of Einstein 
gravity \cite{GergelyND2}.

The metrics (\ref{metric2nd}) are invariant w. r. to the simultaneous
interchange $\left( L,B\right) \rightarrow \left( -L,-B\right) $. This
suggest the existence of two identical copies of the space-time. Cf. 
\cite{GergelyND3} for each value of $B$ there are two copies of
the singularity $R=0$ which devide the range of the coordinate $L$ in three
distinct regions (Fig 1a). From among these $L\in \left( -\infty
,L^{-}<0\right) $ and $L\in \left( L^{+}>0,\infty \right) $ are two
identical copies of the static naked singularity given locally by the metric
(\ref{metric2nd}), case $c=-1.$ The patches $\left( L^{-},0\right) $ and $%
\left( 0,L^{+}\right) $ can be smoothly glued through $L=0$ in order to
obtain a single homogeneous closed Kantowski-Sachs type universe. In this
latter case $R$ is interpreted as the time-dependent radius of the universe
described locally by the metric (\ref{metric2nd}), case $c=1$, which is born
from and collapses to the singularity $R=0$. During its ephemeral existence,
the energy density of the universe evolves cf. Fig 1b.

\begin{figure}[tbh]
\hspace*{.2in}
\special{hscale=30 vscale=30 hoffset=-20.0 voffset=20.0
         angle=-90.0 psfile=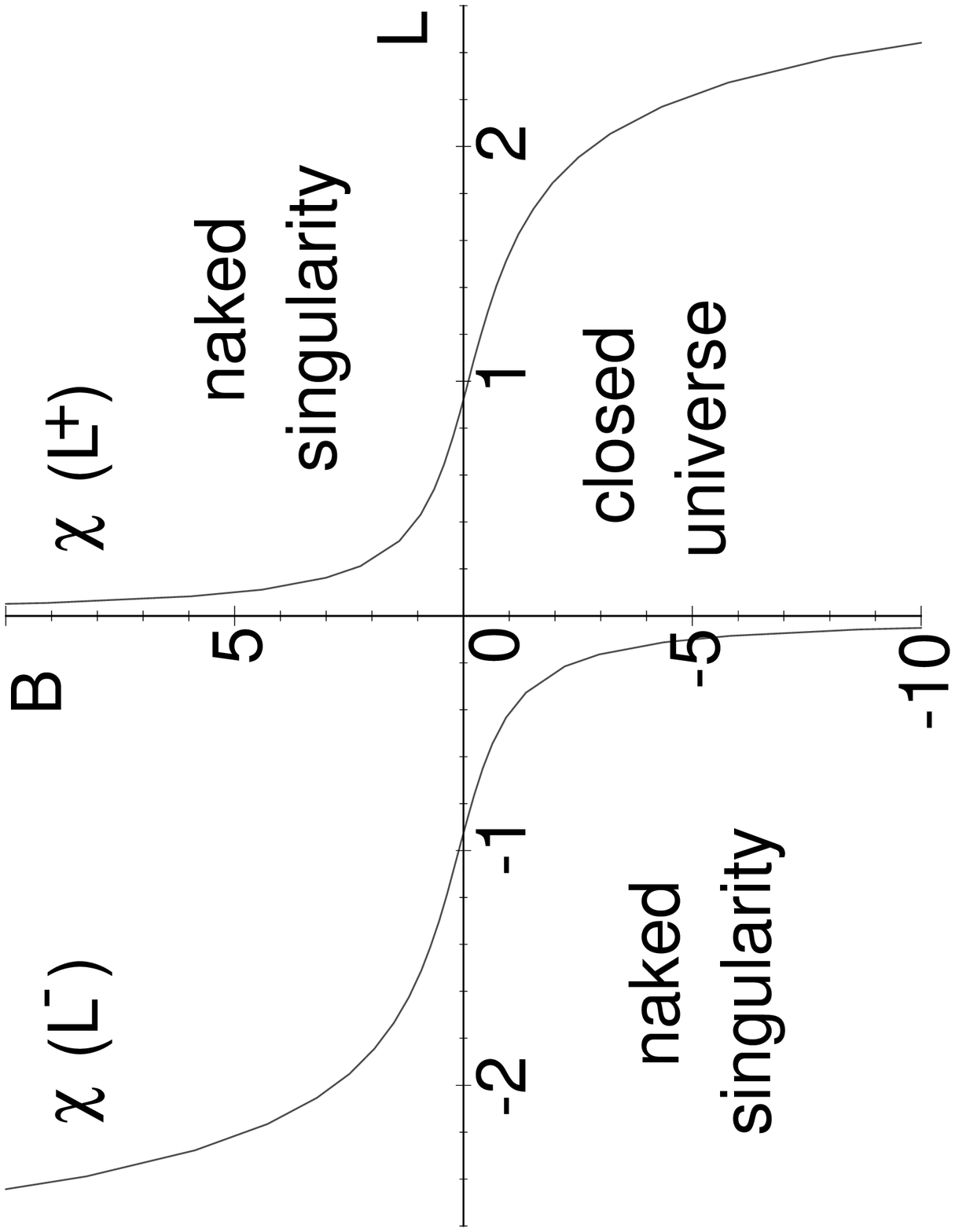}
\hspace*{3.2in}
\special{hscale=30 vscale=30 hoffset=-20.0 voffset=20.0
         angle=-90.0 psfile=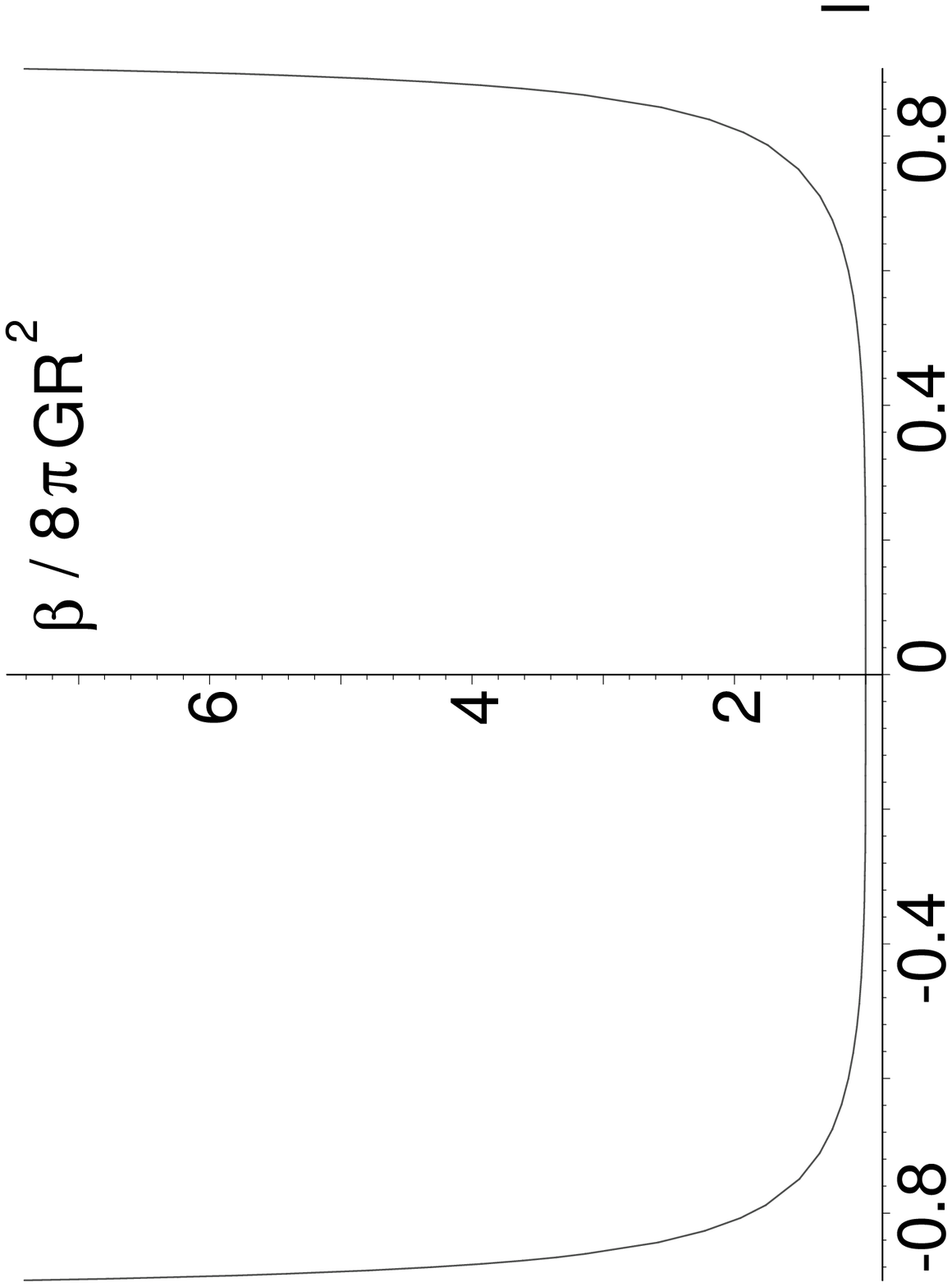}
\vspace*{2.3in}
\caption{ (a) The singularity $R=0$ of the metric (\ref{metric2nd}) is
represented by the curves $B=\protect\chi (L^{\pm})$ which split the range
of the coordinate $L$ in three distinct regions: static, homogeneous,
static. (b) Evolution of the energy density of the null dust universe in
coordinate time $L$ (in units $a=8\pi G$). }
\end{figure}

{\bf Solutions describing the crossflow of ghost fields} can be found by
substituting%
\begin{equation}
L=il\ ,\qquad Z=i\tau \ ,\qquad B=ib\   \label{subst}
\end{equation}%
in the metric (\ref{metric2nd}). Then 
\begin{equation}
\Phi _{B}=i\left( b+\int^{l}e^{-x^{2}}dx\right) =i\phi \ ,\qquad cR=a\left(
e^{-l^{2}}+2l\phi \right) \ .  \label{cRl}
\end{equation}%
and the new metric becomes%
\begin{equation}
ds^{2}=\frac{a}{e^{l^{2}}cR}\left( -d\tau ^{2}+2R^{2}dl^{2}\right)
+R^{2}d\Omega ^{2}\ .  \label{metric2gh}
\end{equation}%
These are the static solutions of Hayward for $c=1$ ($\tau $ time) and new
homogeneous solutions for $c=-1$ ($l$ time). The signature of the metric (%
\ref{metric2gh}) being $\left( -c,c,+,+\right) $, the propagation null
vectors are 
\begin{equation}
\sqrt{2}u_{\pm }=\frac{1}{\sqrt{-cg_{00}}}\frac{\partial }{\partial \tau }%
\pm \frac{1}{\sqrt{cg_{11}}}\frac{\partial }{\partial l}\ .
\end{equation}%
The energy-momentum tensor 
\begin{equation}
T_{ab}=\frac{\beta }{8\pi GR^{2}}\left( u_{a}^{+}u_{b}^{+}\pm
u_{a}^{-}u_{b}^{-}\right) 
\end{equation}%
contains a {\it negative} $\beta $, again irrespective of the value of $c$:%
\begin{equation}
a\beta =-e^{l^{2}}R\ .
\end{equation}%
Therefore the source represents a crossflow of incoming and outgoing ghost
radiation. Alternatively it can be regarded as an anisotropic fluid with
no tangential pressures and both the energy density and radial pressure
equal to $\beta /8\pi GR^{2}=-e^{l^{2}}/8\pi GaR<0$ (infinite at $R=0$). In two 
dimensions it represents a massless scalar field.

As ${\cal R}_{ab}{\cal R}^{ab}=2e^{2l^{2}}/a^{2}R^{2}$ and the Kretschmann
scalar ${\cal R}_{abcd}{\cal R}^{abcd}$ still contains $R^{6}$ in the
denominator, the singular character of the solutions at $R=0$ continues to
hold. The constant $b$ can be related to the value $l_{0\text{ }}$of the
coordinate $l$ at the singularity:%
\begin{equation}
b=\zeta \left( l_{0}\right) =-\frac{1}{2l_{0}e^{l_{0}^{2}}}-\frac{1}{2l_{0}}%
\int^{l_{0}}e^{-x^{2}}dx\ .  \label{bl0}
\end{equation}%
As this equation has no solution for $b\in (-b_{cr},b_{cr})$, with $%
b_{cr}=.8862269255$ (Fig\ 2a.), the singularity is absent for the above
parameter range, and the corresponding solutions are the wormholes of
Hayward. This is the major difference w. r. to the null dust case.

For any other value of $b$ the singularity is present. We discuss next this
possibility. First we remark the symmetry of the metric w. r. to the
simultaneous interchange $\left( l,b\right) \rightarrow \left( -l,-b\right) $%
. In particular the two branches of the curve $\zeta \left( l_{0}\right) $
represent the same singularity. By inspecting Eq. (\ref{cRl}), written in
the form 
\begin{equation}
cR=2al\left[ \zeta \left( l\right) -\zeta \left( l_{0}\right) \right] \ ,
\label{cRzeta}
\end{equation}%
with the remark that $\zeta $ is a monotonously decreasing function ($d\zeta
/dl=-e^{-l^{2}}$) we see that for $b<-b_{cr}$ the singularity devides the $%
l>0$ coordinate range in two domains, characterized by $c=1$ and $c=-1$,
respectively. Similarly for $b>b_{cr}$ the singularity splits $l<0$ in the $%
c=-1,\ c=1$ domains.

\begin{figure}[tbh]
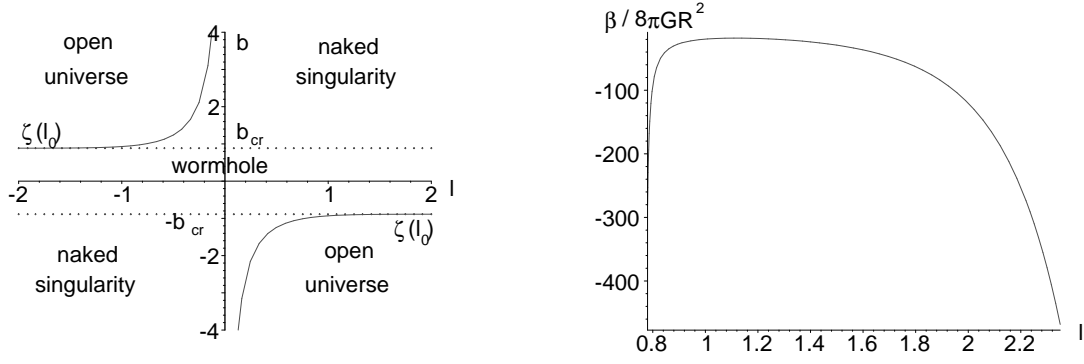

\hspace*{.2in}
\special{hscale=30 vscale=30 hoffset=-20.0 voffset=20.0
         angle=-90.0 psfile=fig2a.eps}
\hspace*{3.2in}
\special{hscale=30 vscale=30 hoffset=-20.0 voffset=20.0
         angle=-90.0 psfile=fig2b.eps}
\vspace*{2.3in}
\caption{ (a) The singularity $R=0$ of the metric \ref{metric2gh} is
represented by the curves $b=\protect\zeta(l_0)$. These split the range
of the coordinate $l$ in two distinct regions for any $b\in
(-\infty,-b_{cr})\cup (b_{cr},\infty)$. For the parameter values $b\in
(-b_{cr},b_{cr})$ there is no singularity. 
(b) Evolution of the energy density of the ghost radiation filled universe in
coordinate time $l$ (in units $a=8\pi G$). }
\end{figure}

{\bf Homogeneous solutions} ($c=-1$ case). These solutions are new and they
lie ''outside'' the two branches of the curve $\zeta \left( l_{0}\right) .$
The evolution of the radius $R$ in coordinate time $l$ is shown on Fig 3.
The solutions represent open universes, either expanding ($b<-b_{cr}$) or
contracting ($b>b_{cr}$). This is to be contrasted with the null dust case,
where the universe was closed. The evolution of the energy density in the
expanding case is shown in Fig 2b. Contrarily with open universes filled
with ordinary matter, here the magnitude of the energy density first
decreases from its original singular value but afterwards it increases again
towards infinity. 

\begin{figure}[tbh]
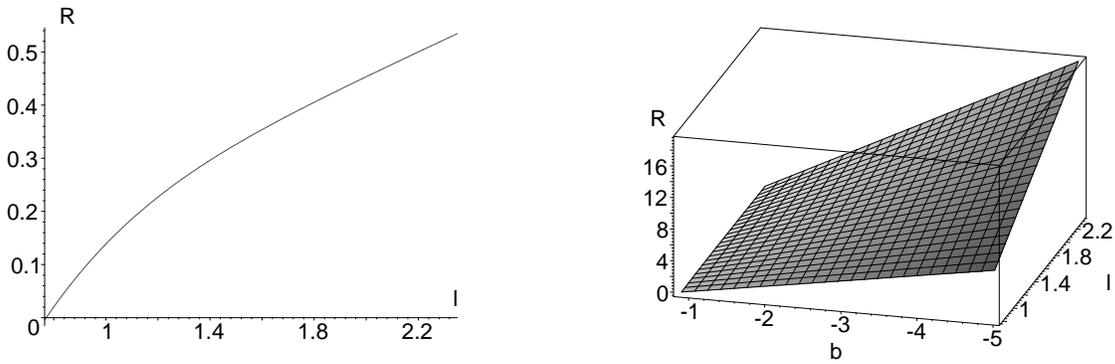

\hspace*{.2in}
\special{hscale=30 vscale=30 hoffset=-20.0 voffset=20.0
         angle=-90.0 psfile=fig3a.eps}
\hspace*{3.2in}
\special{hscale=30 vscale=30 hoffset=-20.0 voffset=20.0
         angle=-90.0 psfile=fig3b.eps}
\vspace*{2.3in}
\caption{ (a) The homogeneous solution represents an open universe (plot for 
$b=-1$).
(b) The universe expands for any admissible value of $b$. }
\end{figure}
 
{\bf Static solutions} ($c=1$ case). The region between the two branches of
the curve $\zeta \left( l_{0}\right) $ contain the static solutions. They
are either wormholes or naked singularities. We verify this latter
possibility by rewriting the metric (\ref{metric2gh}), case $c=1$ as%
\[
ds^{2}=-\frac{a}{e^{-l^{2}}R}dx^{\pm }\left( dx^{\pm }-2\sqrt{2}Rdl\right)
+R^{2}d\Omega ^{2}\ ,
\]
where the null coordinate is given by $dx^{\pm }=d\tau \pm \sqrt{2}Rdl.$
Radial null geodesics are then characterized by%
\begin{equation}
\frac{dl}{dx^{\pm }}=\frac{1}{2\sqrt{2}R}>0  \label{hor}
\end{equation}%
and there is no horizon. 

\begin{figure}[tbh]
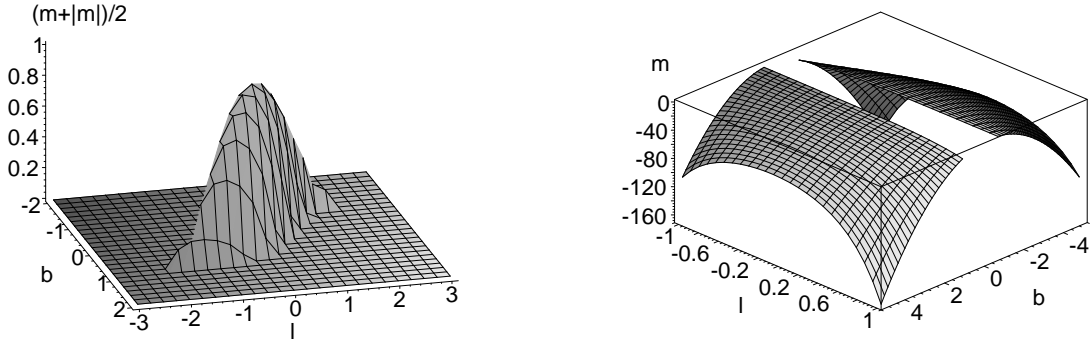

\hspace*{.2in}
\special{hscale=30 vscale=30 hoffset=-20.0 voffset=20.0
         angle=-90.0 psfile=fig4a.eps}
\hspace*{3.2in}
\special{hscale=30 vscale=30 hoffset=-20.0 voffset=20.0
         angle=-90.0 psfile=fig4b.eps}
\vspace*{2.3in}
\caption{ (a) The mass function is positive only for a restricted domain 
of $b$, which represents the wormhole in its throat region. (Mass functions 
are plotted for $a=2$.)
(b) The mass function of naked singularities is negative for any $l$.}
\end{figure}
 
The mass function for the static solutions \cite{Hayward} is $m=\left(
e^{-l^{2}}+2l\phi -2e^{2l^{2}}\phi ^{2}\right) a/2$. There is a
domain of the radial coordinate $l$ with the mass positive, but only for
wormholes (Fig 4a). Naked singularities are characterized by negative $m$,
irrespective of the value of the metric parameters (Fig 4b). The mass
function of static wormholes is represented on Fig. 5. By cutting the
wormhole solution at values $l_{\pm }$ (lying in the positive mass range),
and gluing to spherically symmetric exterior solutions (as described in %
\cite{GergelyND}) we obtain a wormhole composed of a crossflow of
negative-energy radiation, still with positive mass.

\begin{figure}[tbh]
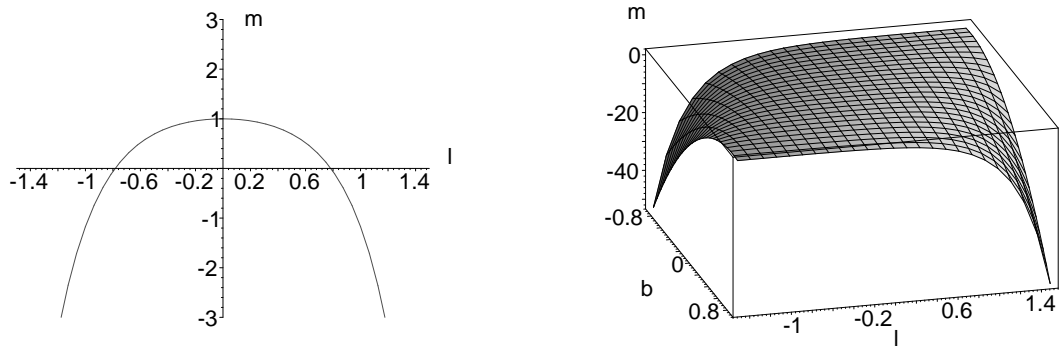

\hspace*{.2in}
\special{hscale=30 vscale=30 hoffset=-20.0 voffset=20.0
         angle=-90.0 psfile=fig5a.eps}
\hspace*{3.2in}
\special{hscale=30 vscale=30 hoffset=-20.0 voffset=20.0
         angle=-90.0 psfile=fig5b.eps}
\vspace*{2.3in}
\caption{ (a) In the throat region the wormhole has positive mass. 
(The plot is done for $b=0$.) 
(b) The missing zone of Fig 4b: the mass function for wormholes, 
represented for various values of the parameter $b$.}
\end{figure}

Summarizing, by switching from a positive to a negative energy density 
(ghost radiation), the static naked singularity is either kept or 
it transmutes to a (possibly positive mass) wormhole. In addition, the 
homogeneous closed universe opens up.

A final remark is that all solutions considered here exhibit a vanishing 
Ricci-scalar, similarly as the wormhole solutions of \cite{Visser}.

{\it Acknowledgments.} This work has been completed under the support of 
the Zolt\'{a}n Magyary Fellowship. Numerical plots were done with MapleV.

\end{document}